# Direct Comparison of Fractional and Integer Quantized Hall Resistance

**Franz J. Ahlers, Martin Götz, and Klaus Pierz**

Physikalisch-Technische Bundesanstalt, Braunschweig, Germany
E-mail: franz.ahlers@ptb.de

**Abstract**
We present precision measurements of the fractional quantized Hall effect where the quantized resistance $R^{[1/3]}$ in the fractional quantum Hall state at filling factor 1/3 was compared with a quantized resistance $R^{[2]}$, represented by an integer quantum Hall state at filling factor 2. A cryogenic current comparator bridge capable of currents down to the nanoampere range was used to directly compare two resistance values of two GaAs-based devices located in two cryostats. A value of 1 - $(5.3 \pm 6.3)$ $10^{-8}$ (95% confidence level) was obtained for the ratio $(R^{[1/3]}/6R^{[2]})$. This constitutes the most precise comparison of integer resistance quantization (in terms of $h/e^2$) in single-particle systems and of fractional quantization in fractionally charged quasi-particle systems. While not relevant for practical metrology, such a test of the validity of the underlying physics is of significance in the context of the upcoming revision of the SI.

## 1. Introduction

The quantized Hall effect, discovered by von Klitzing in 1980 [1], and the previously discovered Josephson effect [2] allow to represent the electrical units ohm and volt in terms of Planck's constant *h* and elementary charge *e*. The effects form the two strongest pillars for the future revised SI [3], since not only the system of electrical units rests on them, but also the unit kilogram, which will in future be realized via those electrical effects by relating virtual mechanical to electrical power in a so-called Kibble balance [4].

Although many results regarding the QHE can be described by disorder phenomena within the edge-state model, its full theoretical description is considerably more involved. Only recently, the developments in the theory of topologically protected states [5] are beginning to provide a unified view of the effect. Therefore, it has been a continuous quest to put the theoretically predicted [6,7] universality of the QHE under experimental challenge by comparing the quantization of resistance in systems which are physically as diverse as possible. Most noteworthy among these are comparisons between 2-dimensional electron systems (2DES) in GaAs/AlGaAs heterostructures and Si-MOSFETs [8,9], and, more recently, between GaAs-based and graphene-based 2DES [10,11]. In these measurements, the quantized Hall resistance (QHR) was of the integer type, involving conventional quasi-particles like electrons in GaAs and Si, or Dirac fermions in graphene, with resistance values predicted by theory as *integer* sub-multiples of $h/e^2$.

Much more exotic quasi-particles are formed, on the other hand, when in very clean systems scattering is suppressed to an extent that many-body interactions of the carriers become dominant. The fractional quantized Hall effect (fQHE), discovered in 1982 [12], is characterized by *fractional* submultiples of $h/e^2$. In a simplified picture it can be understood as an integer QHE (iQHE) of quasi-particles consisting of electrons bound to an even number $2m$ of vortices ('magnetic flux quanta') [13,14]. At filling factors $n/(2mn+1)$ their Hall resistance is given by $(2mn+1)/n$ in units of $h/e^2$. Of all the fractional states, the one with $m, n = 1$ at filling factor 1/3 is the most stable one and was therefore used in our study. The fQHE has been observed in high-mobility GaAs/AlGaAs electron and hole systems [15], as well as in graphene [16]. In semiconductors of extremely high mobility a whole hierarchy of such composite quasi-particles appears [17]. Although it has been discussed whether corrections to exact quantization exist which are specific to the fractional regime [18], the common view is that such corrections are not significant.

Yet, a successful experimental challenge of the universality between fQHE and iQHE systems would constitute one of the strongest supports for the new SI, and it might surprise at first that such a challenge was up to now only performed once [19], at an uncertainty level of 2 parts in $10^6$. The reason for this difficulty lies in the fact that low relative measurement uncertainties of 1 part in $10^9$ or lower can only be achieved at measurement currents of tens of microamperes due to the required low noise level. While such high current levels and the accompanying rise of electron temperature are tolerated by GaAs based QHE devices, and even higher currents by graphene devices [11,20], the quasi-particles responsible for the fQHE are so fragile that they require electron temperatures well below 100 mK to survive.

In this paper, we report the universality test of fQHE and iQHE Hall resistances. The measurements were performed with a cryogenic current comparator (CCC) bridge which was partly rebuilt, especially with respect to low-current operation. We could confirm an agreement with the expected value with a total combined uncertainty of 6 parts in $10^8$ (at 95% confidence level), more than thirty times lower than in [19].

In the following we describe the sample preparation, the key features of the newly built bridge, and discuss the contribution of an additional, but often ignored type-B uncertainty which becomes relevant at low current levels. Finally, we present the measurement data and their detailed analysis and discuss the result.

## 2. Sample preparation

Two GaAs/AlGaAs heterostructure devices were used for the measurements. One of them had a carrier density of $5.0 \cdot 10^{15}$ m$^{-2}$ at a mobility μ of 50 m$^2$/Vs and was used as the iQHE reference device. It had been grown in PTB's standard MBE system and its typical layer sequence GaAs-Al$_{0.3}$Ga$_{0.7}$As-Al$_{0.3}$Ga$_{0.7}$As(Si)-GaAs is shown in the left part of Figure 1(a). This specific device P137-18 has been in use as PTB's standard for high precision calibrations and in international comparisons for several years. Nevertheless, we performed another careful characterization of this device per the *Technical Guidelines* [21] document, right before the universality test described here. The outstanding quality of the device was confirmed, and for the comparison its middle Hall contact pair was used.

The second device was a heterostructure grown in PTB's high-mobility MBE system specifically for these measurements. Its layer sequence was similar, except for a larger spacer thickness (75 nm) between the GaAs-AlGaAs interface and the Si doped layer, and for the fact that Si δ-doping was used instead of volume doping. Also, the thickness from the δ-doping to the capping GaAs layer at the surface was increased to 350 nm. The carrier density of this device was $1.3 \cdot 10^{15}$ m$^{-2}$ and its mobility, measured in the dark at 4 K, was 460 m$^2$/Vs.

A special difficulty with such devices is to obtain low contact resistances. For both the iQHE and the fQHE device we used alloyed Sn-ball contacts, which are known to deliver robust and low-resistance contacts. For the high-mobility fQHE device, however, this is more challenging due to the larger distance between the surface and the 2DEG layer, as schematically illustrated in Figure 1(a). Yet, we achieved contact resistances lower than 10 Ω for all contacts, a value compatible with the requirements for precision measurements [21] in the iQHE regime. As an overview, Figure 1(b) shows the magnetic field dependence of the longitudinal resistance $R_{xx}$ of the fQHE device over a wide magnetic field range. Precision measurements were performed at the center of the plateau at filling factor 1/3 at 16.24 T, indicated in the figure.

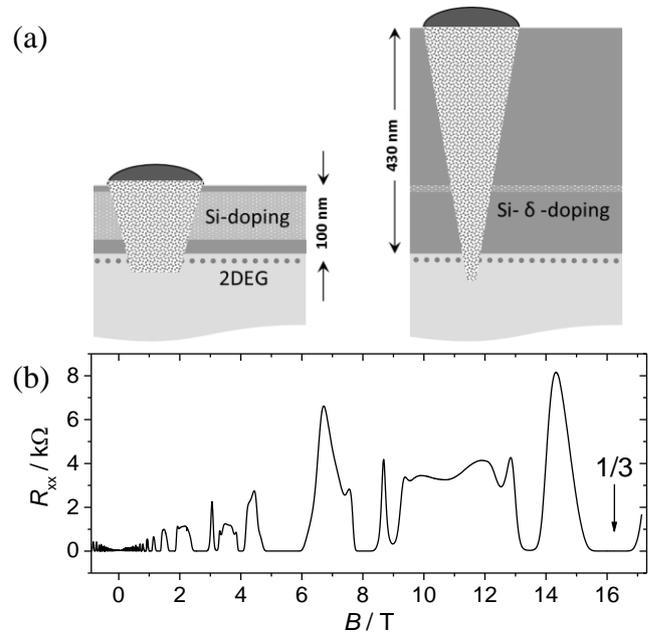

**Figure 1** (a) Layer sequence of the GaAs/AlGaAs iQHR (left) and fQHR heterostructures (right). The 2DES, indicated by dots, is located at the interface between GaAs (light-grey) and AlGaAs (dark grey). The alloyed contacts are indicated by shaded areas. (b) Magnetic field dependence of the longitudinal resistance $R_{xx}$ of the fQHE device, measured with a current of 1 μA at $T = 40$ mK. The position of filling factor 1/3 is indicated.

## 3. Measurement conditions

### 3.1 Bridge setup

The bridge setup comprised two cryo-magnets hosting the iQH and the fQH resistances to be compared. The iQHR cryostat was a standard LHe bath cryostat with a superconducting magnet operated at 10 T (the center of the filling factor 2 plateau of the iQHE device) whose temperature was held at 2.2 K by a λ-cooler. The fQHR cryostat was a top-loading dilution refrigerator equipped with a solenoid capable of 18 T at 4.2 K. A bath temperature of 40 mK was typically used, but depending on current level the electron temperature of the fQHE device was higher. From previously measured temperature dependences of Shubnikov-de Haas oscillations of similar samples it was estimated that a current of 1 µA would cause an electron temperature of around 100 mK under these conditions.

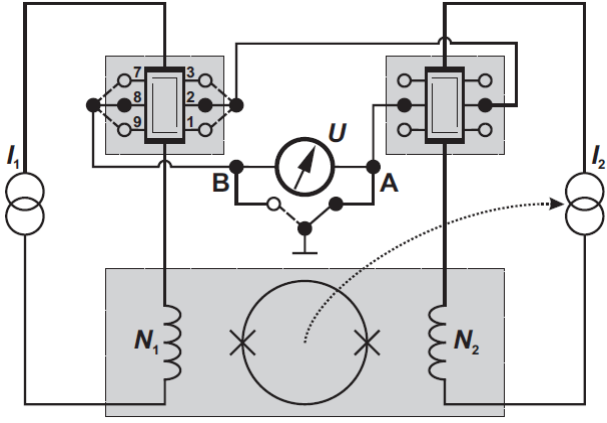

**Figure 2** Scheme of the setup for a direct comparison between fQHR (primary circuit, current source $I_1$) and iQHR (secondary circuit, source $I_2$). Modules operated at low temperature are highlighted by shaded boxes: the fQHR is run in a dilution refrigerator at $T < 50$ mK, the iQHR at about 2.2 K, and the CCC/SQUID probe at 4.2 K. The three cryo-systems were in three rooms. L-shaped lines of equal potential for a given direction of magnetic field are indicated within the Hall bar areas. The numbers of turns $N_1$ and $N_2$ were 3840 and 640, respectively. The dotted arrow is a symbolic representation of the feedback loop which ensured a constant ratio $I_1/I_2 = N_1/N_2 = 6$. Switching the required connection to reference potential between A and B allows to detect whether a leakage path in parallel to one of the Hall bars influences the measurement.

The resistance comparison was performed with a cryogenic current comparator bridge [22] which featured, at the core of its feedback loop, a DC SQUID to detect flux balance of the coils $N_1$ and $N_2$, with oppositely flowing currents, all operated in a Helium dewar at 4.2 K. The nanovolt detector used in the bridge is the one described in [23]. A schematic of the cabling of the measurement is shown in Figure 2. Different from the situation when comparing two standard resistors, or a standard resistor with the iQHR, the use of an auxiliary winding for compensating the deviation from a perfect integer ratio is not needed here. With the choice of a number-of-turns ratio $N_1/N_2$ equal to the 6:1 ratio of fQHR and iQHR, the relative deviation from this ratio is then simply obtained as $\Delta U / \Delta(I_i R_i)$. Here $\Delta U$ is the average bridge voltage difference during synchronous reversals of the currents $I_1$ and $I_2$ through resistor $R_1 = R^{[1/3]}$ (fQHR) and resistor $R_2 = R^{[2]}$ (iQHR), and $\Delta(I_i R_i)$ ($i = 1,2$) represents the voltage drop across each of those. The influence of thermal voltages and their drifts is practically eliminated by the current reversals, which had typical reversal periods of tens of seconds, corresponding to an effective measurement frequency of order tens of millihertz. Transient artefacts due to current reversals were eliminated by discarding the first half of the data points of each reversal half-cycle. The influence of a possible leakage resistance (most harmful when in parallel to the high-resistance fQHR arm of the bridge) was reduced by using shielded and guarded cabling. Nevertheless, it was considered as a type-B uncertainty, assuming a worst-case leakage resistance of 50 TΩ. This value derives from the previously determined $10^{14}$ Ω isolation resistance of our standard QHE-setup, downscaled to take the longer cable path to the fQHE cryostat into account. The corresponding uncertainty $u_L$ is given in line 5 of Table I.

### 3.2 Measurement parameters

Direct comparisons have been performed against the middle Hall contact pair of the iQHR, exposed to a field of 10 T, but varying the Hall contact pairs for the fQHR, to determine longitudinal and Hall resistances. Other variations of experimental parameters include the magnetic field for the fQHR, the settings of the current reversal cycles, and the current bias level (see Figure 3) which was varied from 82 nA to 1.3 µA.

The necessity for low current levels derives from the following consideration: The quasi-particle gap of composite fermions, as determined experimentally in [24,25], is at filling factor 1/3 approximately 0.7 meV (see Fig. 3 in [24], obtained with a sample of very similar carrier density than ours), which is 25 times smaller than the Landau gap of GaAs iQHE devices at 10 T. In addition to limiting the typical iQHR operating temperature of 1.4 K to below 50 mK for fQHE devices, this also sets a limit for the Hall electric field which causes a breakdown of quantization when it becomes too large. The Hall field is proportional to current times resistance, and therefore, due to the six

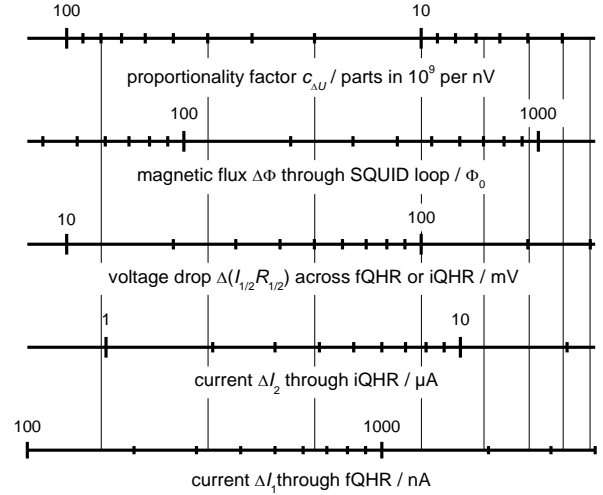

**Figure 3** Scales of the experiments: From bottom to top, the currents flowing through the fQHR or iQHR, respectively, the voltage drop across each of the resistances, and the flux level coupled into the SQUID loop as generated in both the primary and secondary windings are referred to each other. The uppermost scale displays the proportionality factor $c_{\Delta U}$ between the absolute value of $\Delta U$ and the relative deviation of the fQHR-to-iQHR ratio from the expected value of 6. The settings chosen for the experiments are indicated by vertical lines. All units of currents, voltages or flux refer to peak-to-peak values of the current reversal cycles.

times higher resistance, an additional 6-fold decrease of current is required.

While at the typical 40 µA currents used with iQHE devices type-A dominated uncertainties of parts in $10^9$ or lower are routinely achieved with CCC bridges, the sub-microampere current level of the fQHE measurements will not allow such low type-A uncertainties. Also, note that the absolute number of superconducting magnetic flux quanta $\Phi_0$ seen by the SQUID flux balance detector of the CCC bridge becomes as small as ±30 at the lowest current. In consequence, the $1/f$ SQUID noise begins to dominate other noise sources. We reduced this effect by employing a new two-stage SQUID with improved noise figure, as described in detail in [26].

*3.3 Noise rectification*

Further, in this low-current regime, a usually negligible type-B uncertainty contribution cannot be fully ignored any more. It stems from the fact that at low currents and concomitant low flux levels, down-conversion and rectification of high frequency noise at the non-linear SQUID characteristic becomes more and more significant and can systematically falsify the reading of the bridge. This effect, described in detail in [27], would require very long averaging times to quantify it precisely, with no guarantee that at the actual resistance measurement the same conditions prevail. For our results presented in the next section we estimated as an upper limit for its influence a flux error of ±1 µ$\Phi_0$. This is treated as the limit of a rectangular distribution and represented as a type-B uncertainty in line 4 of Table I.

At the lowest current level the influence is strongest, yielding an absolute uncertainty $3R_K(1\mu\Phi_0/29.5\Phi_0)$ of 2.63 milliohm. This additional uncertainty increases the total uncertainty by 40% at this current (and less at higher currents), which does not yet limit the significance of our results severely. However, should resistance comparisons at even lower currents be attempted, e.g. when testing the precision of quantization of the recently demonstrated quantized anomalous Hall effect of ferromagnetic 3-dimensional topological insulators [28-30], this contribution must be considered.

*3.4 Determination of resistances*

Because the bridge can only be balanced when it measures a Hall voltage, the longitudinal voltages were obtained as differences of Hall voltages, as is recommended practice in precision resistance measurements (see section 6.2 in [21]). For each single set of measurements at a given magnetic field and current, four Hall voltages $a, b, c, d$ were determined. They are indicated by the "bow-tie" arrow pattern in Figure 4. From these voltages and the known current, longitudinal resistances and differences of resistances to the reference resistance can be calculated, as described in [21] and below, provided the consistency condition $a - c + d - b = 0$ is fulfilled. Unlike recommended in [21], however, we refrained from averaging all four voltages to obtain a Hall voltage with lower type-A uncertainty. We instead restricted ourselves to calculating $\delta R_{xy} := R^{[1/3]} - 6R^{[2]}$ as $(a+d)/2I$ and $R_{xx}$ as $(b-c)/2I$ (writing from here on just $I$ for the fQHE current instead of $I_1$ as in Figure 2). This way correlations between the $\delta R_{xy}$ and $R_{xx}$ data are avoided and statistics subtleties in the subsequent analysis of the $\delta R_{xy}(R_{xx})$ dependence need not be considered.

Note that the low-resistance, but comparatively large Sn-ball contacts cause a small longitudinal contribution to the measured Hall resistance even for geometrically exactly opposing contacts [31], leading to an apparent linear contribution of $R_{xx}$ to $\delta R_{xy}$. Estimated from the contact and Hall bar widths, this alone would contribute approx. 8% of $R_{xx}$ to the measured $\delta R_{xy}$ values, as symbolized by the slightly tilted arrows $a$ and $d$.

It is known [32-35] that in iQHE devices also thermally activated transport contributes to the linear correlation between $\delta R_{xy}$ and $R_{xx}$, and often dominates it. Such effects are likely to occur in an fQHE device already at the sub-µA current levels used here, due to the smaller energy gap and the fragility of the fractional state. Therefore, one must rely on an extrapolation of the $\delta R_{xy}(R_{xx})$ dependence to zero $R_{xx}$ for the determination of the 'true' $\delta R_{xy}$ value. The demonstration that the extrapolated $R_{xx}(I)$ becomes zero for zero current is of course a prerequisite for this.

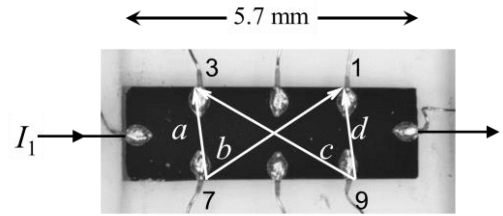

**Figure 4** Photo of the fQHE Hall bar (dark rectangle) with Sn-ball contacts. Current flows through the left and right contacts. Following Figure 2, the voltage probe contacts are labelled 1 and 3 on the high potential side and 9 and 7 on the low potential side. White arrows indicate the measured bridge voltages $a$ to $d$ from which longitudinal and Hall resistances are obtained.

## 4. Results

*4.1 Summary of measured data*

As a preparation step for the comparison we determined the center of the plateau at filling factor 1/3 from a series of measurements with 1 µA current at six different magnetic fields between 16.1 and 16.4 T. From the six voltage data sets $\{a, b, c, d\}$, longitudinal resistance values were obtained as described above, and the minimum of a parabola fitted to those data determined the magnetic field position of 16.24 T where the actual comparison was performed. Next, six more data sets $\{a, b, c, d\}$ were measured at this field with six different current levels ranging from 0.08 to 1.3 µA. The measurement time for one data set was 88 minutes, except for the two lowest currents, where it was four times longer, nearly 6 hours per set.

The results are summarized in Table I, which lists current values and measurement durations in lines 1 and 2. Line 3 lists the flux seen by the SQUID, for the given current level and the number of windings given in the caption of Figure 2. In lines 4 and 5 the type-B uncertainties described in sections 3.3 and 3.2 are listed. The actual bridge voltage readings $a$ to $d$ and their type-A uncertainties, as obtained from the current reversal cycles, are listed in lines 6 to 9, with the consistency check value $a - c + d - b$ in line 10.

Finally, the last two lines give the longitudinal resistance $R_{xx}$ and the resistance difference $\delta R_{xy}$. The uncertainties in lines 6 to 10 result from the type-A uncertainties of the $\{a,b,c,d\}$ values only, whereas for the bold values the type-B uncertainties from lines 4 and 5 have been included, assuming a rectangular probability distribution for the type-B components.

**Table I** Measurement currents $I$, measurement durations $t$ for one voltage set $\{a,b,c,d\}$, maximum flux levels seen by the SQUID, and estimated type-B uncertainties $u_B$ for the given flux level (Lines 1 to 5). Actual bridge voltage readings $a$ to $d$ and the consistency check value $a - c + d - b$, with type-A uncertainties (67% confidence level) obtained from the statistical distribution of the raw data values, are in Lines 6 to 10. The last two lines list longitudinal resistances $R_{xx} = (b-c)/2I$ and resistance deviations $\delta R_{xy} = (a+d)/2I$, both in milliohm. Their uncertainties include $u_B$ and $u_L$ from lines 4 and 5 and were calculated as $\sqrt{u_A^2 + u_B^2/3 + u_L^2/3}$, with $u_A$ calculated from the $\{a,b,c,d\}$ uncertainties.

| $I$ / μA | ±0.0820 | ±0.1615 | ±0.3260 | ±0.6520 | ±0.9735 | ±1.3025 |
|---|---|---|---|---|---|---|
| $t$ / min | 352 | 352 | 88 | 88 | 88 | 88 |
| Flux / $\Phi_0$ | ±29.5 | ±58.2 | ±117 | ±235 | ±351 | ±469 |
| $u_B$ / mΩ | 2.63 | 1.33 | 0.67 | 0.33 | 0.22 | 0.17 |
| $u_L$ / mΩ | 0.12 | 0.12 | 0.12 | 0.12 | 0.12 | 0.12 |
| $a$ / nV | -0.70 ± 0.35 | -0.65 ± 0.27 | -4.21 ± 0.50 | -21.09 ± 0.69 | -53.97 ± 0.73 | -151.54 ± 0.54 |
| $b$ / nV | -0.66 ± 0.38 | -0.61 ± 0.31 | -2.16 ± 0.53 | -7.88 ± 0.83 | 7.60 ± 0.59 | 23.09 ± 0.41 |
| $c$ / nV | -0.27 ± 0.22 | -0.85 ± 0.28 | -4.75 ± 0.45 | -26.38 ± 0.81 | -97.54 ± 1.04 | -266.41 ± 0.45 |
| $d$ / nV | -0.48 ± 0.37 | -0.74 ± 0.25 | -2.50 ± 0.55 | -12.87 ± 0.65 | -34.06 ± 0.76 | -93.01 ± 0.56 |
| $a - c + d - b$ / nV | -0.25 ± 0.67 | 0.07 ± 0.56 | 0.19 ± 1.01 | 0.30 ± 1.50 | 1.90 ± 1.59 | -1.23 ± 0.99 |
| $R_{xx}$ / mΩ | **-1.19 ± 2.03** | **0.37 ± 1.01** | **1.99 ± 0.66** | **7.09 ± 0.48** | **27.00 ± 0.34** | **55.57 ± 0.17** |
| $\delta R_{xy}$ / mΩ | **-3.59 ± 2.17** | **-2.16 ± 0.96** | **-5.15 ± 0.69** | **-13.02 ± 0.41** | **-22.61 ± 0.31** | **-46,94 ± 0.20** |

*4.2 Extrapolation to zero current*

Since the longitudinal resistance is zero only at the lowest current levels where the relative measurement uncertainty is rather high, we use an extrapolation of $R_{xx}$ and $\delta R_{xy}$ to zero current to obtain more reliable values for these quantities. The graphical representations of the data in bold in Figure 5 and Figure 6 serve to illustrate the extrapolation procedure. For such an extrapolation, it is more important to describe the data by a smooth functional form with as few as possible fit parameters than to find a representation backed by a physical transport model. We used two methods for the extrapolation. In the first method, the current dependences $R_{xx}(I)$ and $\delta R_{xy}(I)$ are each described by a simple power law, and the value of this function at $I = 0$ is taken as an estimate of the extrapolated value. This is shown in Figure 5. For the second method, the data are plotted as $\delta R_{xy}(R_{xx})$, and in a similar way an extrapolated value $\delta R_{xy}(R_{xx} \rightarrow 0)$ is determined. This is shown in Figure 6.

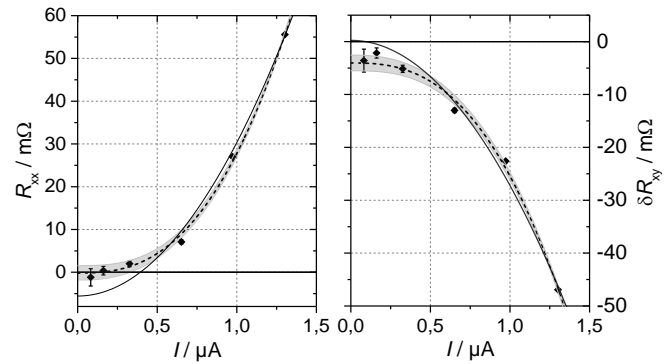

**Figure 5** Plots of $R_{xx}(I)$ and $\delta R_{xy}(I)$ data from Table I. Dashed lines represent two weighted least squares fits of the form $\alpha + \beta I^\gamma$, with the shaded areas indicating the 67% confidence band of the fit. For both curves $\gamma$ was the same, with a value of 2.62 ± 0.18 resulting from the fit. The solid curves represent fits with $\gamma = 2$.

There is no theory predicting how $\delta R_{xy}$ and $R_{xx}$ should depend on current. Empirically, a simple power law as we use here has been observed before, e.g. in [36] for the case of an GaAs iQHE device. If only thermal effects were responsible for the inter-edge-channel scattering and the concomitant rise of $R_{xx}$, it would be tempting to assume an $I^2$ behavior for extrapolating to zero current [37]. However, an attempt to fit an $I^2$-law to our data fails, as the solid lines in Figure 5 show. (Also the data in [36], when closely analyzed, follow an $I^3$ rather than an $I^2$ law). However, since the Hall electric field, which also can induce scattering, is proportional to current, exponents larger than 2 are quite reasonable. Consequently, we chose to perform the weighted least-squares regression analysis of our data with functions of the more general form $\alpha_j + \beta_j I^{\gamma_j}$, with $j = x, y$ for $R_{xx}(I)$ and $\delta R_{xy}(I)$, respectively.

When analyzing the $R_{xx}$ and $\delta R_{xy}$ data sets, the resulting $\gamma_j$ exponents were found identical within their standard error. Therefore, an additional regression was performed with $\gamma$ restricted to be identical for both data sets (reducing the number of fit parameters from 6 to 5 as a side effect). The resulting curves and their 67% confidence bands are shown in Figure 5 as dashed lines and shaded areas, respectively.

The extrapolated value $\alpha_x$ for $R_{xx}(I \to 0)$, together with its 67% and 95% confidence limits, is given in line 2 of Table II. It confirms that at vanishing current the fQHR device is well quantized at filling factor 1/3. The extrapolated value for $\delta R_{xy}(R_{xx}(I) \to 0)$ and its confidence limits are given in line 3. They were obtained by using the fact that, since $R_{xx}$ and $\delta R_{xy}$ are linear in $I^\gamma$, $\delta R_{xy}(I)$ can be written as $\alpha_y + \beta_y(R_{xx}(I) - \alpha_x)/\beta_x$. This gives, in the limit $R_{xx}(I) \to 0$, for $\delta R_{xy}$ the estimate $\alpha_y - \beta_y \alpha_x / \beta_x$ for $\delta R_{xy}(R_{xx}(I) \to 0)$. The confidence limits of this aggregate value were calculated from the confidence limits of the individual $\alpha_j, \beta_j$ fit results.

In the second method of extrapolation, a regression analysis is performed directly on the $\delta R_{xy}(R_{xx})$ data shown in Figure 6, taking errors in both the x- and y-axes into account. The data show a linear trend, as is quite commonly observed in iQHR measurements [21], and as is compatible with the result from method 1. We used York's method [38] for the weighted linear least squares fit. Regarding the confidence estimates, York's algorithm evaluates them at the least-squares-adjusted points rather than at the observed points, thereby producing the same estimates as a maximum likelihood approach. The resulting regression curve and its 67% confidence band are shown in the figure, and the obtained values for the axis intercept and its 67% and 95% confidence limits are given in line 4 of Table II.

Converting the numbers with the 95% uncertainty estimate from method 2 from milliohms to relative values,

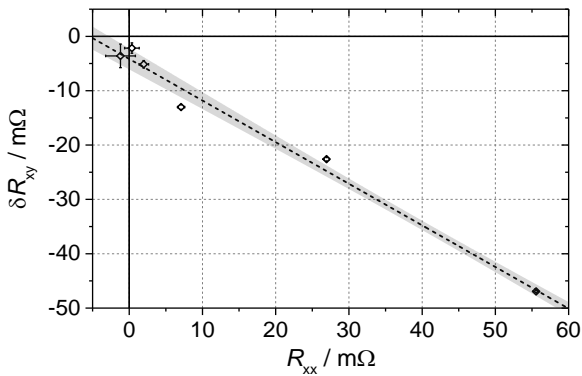

**Figure 6** Plot of $\delta R_{xy}(R_{xx})$ data from Table I. A weighted linear least squares fit with errors in $\delta R_{xy}$ and $R_{xx}$ considered was performed using the York method [38]. The shaded area indicates the 67% confidence band of the fit.

we get as the result of our analysis:

$$R^{[1/3]}/6R^{[2]} = 1 - (5.3 \pm 6.3) \cdot 10^{-8}$$

**Table II** Extrapolated values of $R_{xx}$ and $\delta R_{xy}$ for the limit of vanishing current, assuming power law current dependences $\alpha + \beta I^\gamma$, are given in lines 2 and 3. The extrapolated value of $\delta R_{xy}$ for the limit of vanishing $R_{xx}$, obtained by directly analysing the data set $\delta R_{xy}(R_{xx})$, is given in line 4.

|  | value | 67% confidence | 95% confidence |
|---|---|---|---|
| $R_{xx}(I \to 0)$ | - 0.2 mΩ | ± 1.7 mΩ | ± 3.9 mΩ |
| $\delta R_{xy}(R_{xx}(I) \to 0)$ | - 4.2 mΩ | ± 2.0 mΩ | ± 4.5 mΩ |
| $\delta R_{xy}(R_{xx} \to 0)$ | **- 4.1 mΩ** | **± 1.9 mΩ** | **± 4.9 mΩ** |

*4.3 Discussion*

The universality test presented in this paper can be formally qualified as 'passed', since the expected value of zero is just covered by the 95% confidence limit.

However, we like to point out that in both methods 1 and 2 the goodness-of-fit parameter $\chi^2_{reduced}$ in the numerical regressions was around 9, and not of order unity, as would be expected for correctly chosen models and realistic uncertainty estimates of the measured data [38]. The scatter of our data is obviously larger than the assigned uncertainties, indicating that it is our uncertainty estimate which is not realistic, although all relevant components have been included to the best of our knowledge.

Algorithms performing weighted linear-least-squares fits of data with errors take care of this by scaling the confidence limits by $\sqrt{\chi^2_{reduced}}$ (which is equivalent to scale the uncertainties of the individual data points by the same factor). Known as the 'Birge ratio method' (see e.g. [39]), this is basically an uncertainty estimation based on the experimentally found deviation of the data from the linear adjustment. We believe that the application of this method does not invalidate the result of our regression analysis, but in similar future studies advanced regression methods should be applied, when appropriate tools become more widely available. Such methods will likely be based on Bayesian inference, as e.g. described in [40] for the case of linear regression of data with negligible uncertainties in $x$.

## 5. Outlook

The relative measurement uncertainty level of $6.3 \cdot 10^{-8}$ constitutes a record for this kind of measurement at current levels in the nanoampere regime. Yet, and although the result of this universality test qualifies it as 'passed', an even lower uncertainty seems desirable. An obvious reason is that the data leave some room for speculations whether a failure of universality might be observed when measurement uncertainty is reduced further, and, more importantly, what the physics behind such a deviation could be. The only way to end such speculation is indeed a measurement with lower uncertainty.

Another reason to strive for better measurement uncertainty in the low-current regime is the recently discovered quantized anomalous Hall effect in 3-dimensional topological insulators [28 – 30] which requires even lower current levels, at least at the current stage of material development.

We see several routes to an improvement of the uncertainty. One is a more compact arrangement of the resistances to be compared, to minimize noise pickup by long cables. Although we have extremely carefully optimized our experiment in this respect, a setup with shorter cables would be advantageous.

Another route is to increase the flux level seen by the SQUID by increasing the overall number of turns in the CCC by some factor. This will, firstly, relax the down-mixing effects leading to the type-B uncertainty discussed in section 3.3 by the same factor.

As a second effect, a higher number of turns may reduce the contribution of the intrinsic SQUID flux noise to the combined type-A uncertainty of the bridge readings $\Delta U$. The reason is that unlike other noise components (thermal noise of the resistors, amplifier noise), the SQUID contribution scales inversely with the number of turns due to the conversion from flux noise to detected voltage noise. The benefit from this ends of course when the SQUID contribution is decreased below the other noise components. For our specific set-up, we estimated that already a 3-times increase of the number of turns would reduce the SQUID noise's influence to insignificance.

We have in the meantime set up a new CCC with an about 4-times higher number of turns [41]. Additionally, in the new hardware electric interference is reduced and extended measurement times are possible, which both is helpful especially with respect to the mentioned parasitic effects. A more precise test of the universality of the QHE also in the fractional regime should thus be possible in future.

## Acknowledgement

The authors acknowledge valuable support by M. Busse, E. Pesel, G. Muchow, H. Marx, and B. Hacke as well as stimulating discussions with Dietmar Drung, Hansjörg Scherer and Katy Klauenberg.